\documentclass[twocolumn,showpacs,aps,prl,superscriptaddress,floatfix]{revtex4}
\usepackage{graphicx}
\usepackage{verbatim}
\usepackage{dcolumn}
\usepackage{amsmath}
\usepackage{epsfig}
\usepackage{subfigure}

\newcommand{\BaBarYear}      {11}
\newcommand{\BaBarNumber}    {025}
\newcommand{\BaBarType}      {PUB}  
\newcommand{\SLACPubNumber}  {14863}

\input babarsym.tex

\def\epem{e^+e^-}
\def\mpmm{\mu^+\mu^-}
\def\lplm{l^+l^-}

\def\pipm{\pi^+\pi^-}

\def\figurebox#1#2#3{%
    \def\arg{#3}%
    \ifx\arg\empty
    {\hfill\vbox{\hsize#2\hrule\hbox to #2{\vrule\hfill\vbox to #1{\hsize#2\vfill}\vrule}\hrule}\hfill}%
    \else
    {\hfill\epsfbox{#3}\hfill}%
    \fi}

\begin{document}

\pagestyle{plain}

\begin{flushleft}
\babar-\BaBarType-\BaBarYear/\BaBarNumber \\
SLAC-PUB-\SLACPubNumber\\
arXiv:1202.1313 [hep-ex]\\
\end{flushleft}

\title{{\large \bf Search for low-mass dark-sector Higgs bosons}}

%
\author{J.~P.~Lees}
\author{V.~Poireau}
\author{V.~Tisserand}
\affiliation{Laboratoire d'Annecy-le-Vieux de Physique des Particules (LAPP), Universit\'e de Savoie, CNRS/IN2P3,  F-74941 Annecy-Le-Vieux, France}
\author{J.~Garra~Tico}
\author{E.~Grauges}
\affiliation{Universitat de Barcelona, Facultat de Fisica, Departament ECM, E-08028 Barcelona, Spain }
\author{D.~A.~Milanes$^{a}$}
\author{A.~Palano$^{ab}$ }
\author{M.~Pappagallo$^{ab}$ }
\affiliation{INFN Sezione di Bari$^{a}$; Dipartimento di Fisica, Universit\`a di Bari$^{b}$, I-70126 Bari, Italy }
\author{G.~Eigen}
\author{B.~Stugu}
\affiliation{University of Bergen, Institute of Physics, N-5007 Bergen, Norway }
% ...
\collaboration{The \babar\ Collaboration}
\noaffiliation

\begin{abstract}

Recent astrophysical and terrestrial experiments have motivated the proposal of a dark sector 
with $\gev$-scale gauge boson force carriers and new Higgs bosons. We present a search for a 
dark Higgs boson using 516 $\rm fb^{-1}$ of data collected with the \babar~ detector. We do 
not observe a significant signal and we set 90\% confidence level upper limits on the product 
of the Standard Model-dark sector mixing angle and the dark sector coupling constant. 
\end{abstract}


\maketitle

\setcounter{footnote}{0}

While the astrophysical evidence for dark matter is now overwhelming, its precise nature 
and origin remain elusive. Recent results from terrestrial and satellite experiments 
have motivated the proposal of a new, hidden gauge sector under which WIMP-like dark matter 
particles are charged \cite{Fayet,ArkaniHamed:2008qn,Pospelov:2007mp}. An Abelian gauge 
field, the dark photon $A'$, couples this dark sector to Standard Model (SM) particles 
through its kinetic mixing with the SM hypercharge fields \cite{Holdom}. In this framework  
dark matter particles can annihilate into pairs of dark photons, which subsequently decay to SM 
particles. The dark photon mass is constrained to be at most a few $\gev$ to be compatible 
with astrophysical constraints \cite{PAMELA,FERMI}. In a minimal model \cite{Batell:2009yf}, 
the dark photon mass is generated via the Higgs mechanism, adding a dark Higgs boson $h'$
to the theory. The mass hierarchy between these two particles is not constrained, and the 
dark Higgs boson could be light as well.

A consequence of this scenario is the possibility of probing a light dark sector at 
low-energy $\epem$ colliders \cite{Batell:2009yf,Essig:2009nc} and fixed-target 
experiments \cite{Bjorken:2009mm,Batell:2009di}. Searches for dark photon production 
have yielded negative results, and constraints have been derived on the mixing strength 
between the SM and the dark sector, $\epsilon$, as a function of the dark photon mass \cite{Bjorken:2009mm}.
  
The Higgs-strahlung process, $\epem \rightarrow A' h', h' \rightarrow A' A'$, 
might offer another gateway to a dark sector. This reaction is of particular interest, since it is 
one of the few process suppressed by a single factor of $\epsilon$, and the background is 
expected to be small. If observed, this reaction could provide an unambiguous signature of 
physics beyond the Standard Model. 
The event topology depends on the dark Higgs and dark photon masses. While Higgs bosons heavier 
than two dark photons decay promptly, their lifetime becomes large enough to escape undetected 
for $m_{h'} < m_{A'}$. Moreover, the dark photon width is proportional to $m_{A'} \epsilon^2$, 
and its decay can be prompt or displaced, depending on the value of these parameters. At 
\babar~ energies, the decay length in the detector is ${\cal O}(100) \rm~\mum$ or less for 
$m_{A'} > 250 \mev$ and $\epsilon \gtrsim 10^{-4}$, and dark photon decays can be considered 
as prompt in this regime.

We report a search for dark Higgs production in the Higgs-strahlung process. The 
measurement is performed in the range $0.8 < m_{h'} < 10.0 \gev$ and $0.25 < m_{A'} < 3.0 \gev$ 
with the constraint $m_{h'} > 2 m_{A'}$. 
To avoid any experimental bias, the data are not examined before the selection 
procedure is finalized. The data sample used in this analysis consists of 521 fb$^{-1}$ of data 
collected mostly at the $\Upsilon(4S)$ resonance, but also including luminosity at the 
$\Upsilon(3S)$ and $\Upsilon(2S)$ peaks, as well as off-resonance data. A sample corresponding 
to $\sim 10\%$ of the data (optimization sample) is used to optimize the selection criteria and 
is discarded from the final dataset. This sample is treated entirely as background for optimization 
and background studies.

The \babar\ detector is described in detail elsewhere \cite{Bib:Babar}. Charged particle momenta 
are measured in a tracking system formed by a five-layer double-sided silicon vertex detector and a 
40-layer central drift chamber both immersed in a 1.5 T axial magnetic field. Electron and 
photon energies are measured in a CsI(Tl) electromagnetic calorimeter. 
Charged-particle identification (PID) is performed using an internally reflecting 
ring-imaging Cherenkov detector and the energy loss $dE/dx$ measured by the silicon 
vertex detector and central drift chamber. Muons are mainly identified by the 
instrumented magnetic flux return. 

Signal events are generated by MadGraph \cite{Alwall:2007st} for 
about 40 different hypotheses of dark photon and Higgs boson masses. The hadronization of the 
$A' \rightarrow q \bar{q}$ ($q=u,d,s,c)$ decay is performed by JETSET \cite{Bib::jetset}. The 
detector acceptance is studied using Monte Carlo (MC) simulation based on GEANT4 \cite{Bib::Geant}. 
Time-dependent detector inefficiencies, as monitored during data taking periods, are included 
in the simulation.

The $\epem \rightarrow  A' h', h' \rightarrow A' A'$ reaction is 
either fully reconstructed in the $3(\lplm)$, $2(\lplm)\pipm$ and $\lplm 2(\pipm)$ final 
states ($l=e,\mu$), or partially reconstructed in the $2(\mpmm)+X$ and $\mpmm \epem +X$ 
channels, where $X$ denotes any final state other than a pair of pions or leptons. The 
$2(\epem) + X$ mode suffers from significantly more background than the other channels and
is excluded. The first modes are collectively referred to as ``exclusive modes'', as opposed 
to ``inclusive modes'' for the $2(\lplm) + X$ channels. The inclusive modes are only 
considered in the region $m_{A'} > 1.2 \gev$, since their contribution is small below this 
threshold and the background level becomes large.

The event selection proceeds by first reconstructing dark photon candidates from pairs of 
oppositely-charged tracks identified as electrons, muons or pions by PID algorithms. 
In addition, the helicity angle of the electron in the dark photon rest frame, $\alpha_e$, must 
satisfy $\cos\alpha_e < 0.9$. The background from accidental $\epem$ pairs exhibits a peaking 
component near $\cos\alpha_e\sim1$, while signal events are broadly distributed.
Events are then processed according to the following sequence of 
hypotheses until a match is found:
$6\mu, 4\mu 2e, 2\mu 4e, 6e, 4\mu 2\pi, 2\mu 2e 2\pi, 4e 2\pi, 2\mu 4\pi, 2e 4\pi, 4\mu + X, 2\mu 2e + X$. 
This order is chosen to minimize the cross-feed between channels and the efficiency loss due to 
misclassification. 

\begin{figure}[!htb]
\begin{center}
\includegraphics[width=0.5\textwidth]{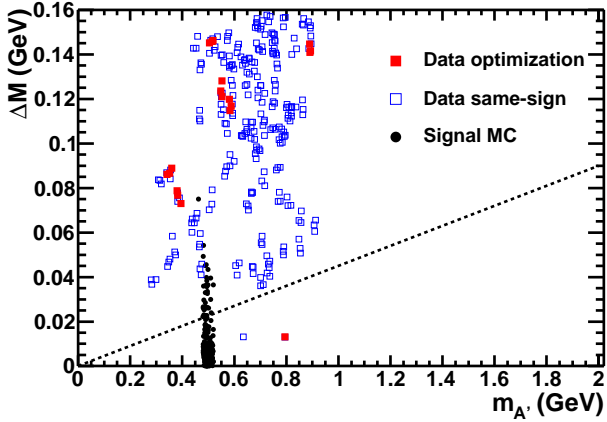}
\caption{Distribution of the largest mass difference between the three dark photon candidates ($\Delta M$) versus 
the average dark photon mass ($m_{A'}$) after all other selection criteria are applied for the $2e 4\pi$ final 
state. The data are shown for opposite-sign combinations from the optimization sample (plain squares) as well as 
an additional background estimation, described later, of same-sign combinations from the full dataset 
(open squares). The Monte Carlo predictions for $m_{h'}=3.0 \gev$ and $m_{A'} = 0.5 \gev$ are displayed 
as plain circles. The signal region for the $2e 4\pi$ mode is delimited by the
dashed line.}
\label{figsel}
\end{center}
\end{figure}

Additional criteria are applied to increase the purity of the signal. Exclusive modes must 
contain exactly six charged tracks, and the invariant mass of the three dark photon system must 
be larger than 95\% of the $\epem$ center-of-mass energy. The dark photons are then fitted, 
constraining the tracks to originate 
from the interaction point. The fit probability is required to be larger than $10^{-5}$. Finally, 
the largest mass difference between the dark photon candidates, $\Delta M$, must be less than 
$10-240 \mev$, depending on the final state and the dark photon masses. The distribution of this variable after 
all other selection criteria are applied is displayed in Fig.~\ref{figsel} for the $2e 4\pi$ final 
state. The signal peaks near $\Delta M \sim 0$, while the background is concentrated towards 
higher values.

Inclusive modes are selected by requiring two leptonic dark photon candidates with similar masses. The two 
dark photons are fitted, constraining the four leptons to originate from the interaction point. Events 
with a fit probability less than $10^{-5}$ are discarded. The remaining dark photon is then identified 
as the system recoiling against the two lepton pairs. The cosine of its polar angle in the laboratory frame 
must be less than 0.99 to remove radiative QED events. Finally, the masses of all dark photons must 
be compatible within their uncertainties. 

\begin{figure}[!htb]
\begin{center}
\includegraphics[width=0.5\textwidth]{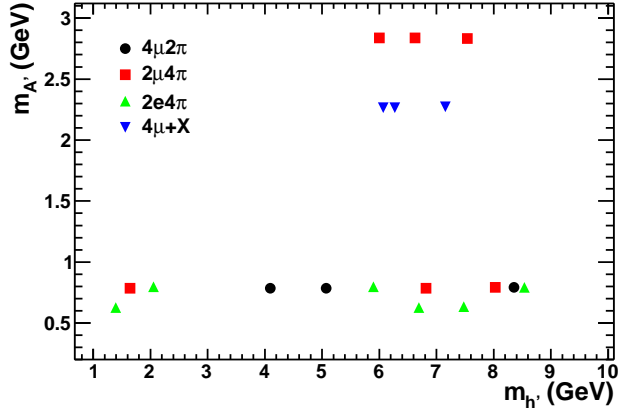}
\caption{Distribution of dark photon mass ($m_{A'}$) versus the dark Higgs mass ($m_{h'}$) for the final data 
sample. Three entries are plotted for each event, corresponding to the possible assignments of the decay 
$h' \rightarrow A' A'$.}
\label{figdat}
\end{center}
\end{figure}

A total of six events are selected by these criteria: one $4\mu 2\pi$, two $2\mu 4\pi$, two $2e 4\pi$ 
and one $4\mu + X$ events. No candidate containing six leptons survives the selection. The distribution of 
the dark photon mass versus the dark Higgs boson mass is shown in Fig.~\ref{figdat}. Three entries, 
corresponding to the possible assignments of the decay $h' \rightarrow A' A'$, are considered for 
each event. Besides the contribution of 
$\rho \rightarrow \pipm$ or $\omega \rightarrow \pipm$ decays near $m_{A'} \sim 0.7-0.8 \gev$, 
no significant signal is observed. This result is consistent with the two events observed in 
the optimization sample, assumed to be background.
Given these limited statistics, a second background estimation based on the full dataset using same-sign 
combinations, such as $(\epem) (\mu^+\mu^+) (\mu^-\mu^-)$ or $(e^+e^+) (\mu^-\mu^-) X$, is 
used as a cross-check. Both methods predict background levels consistent within their 
statistical uncertainties.

Using uniform priors in the cross-section, 90\% confidence level (CL) Bayesian upper limits on the production cross-section are 
derived for each mode separately as a function of the dark Higgs and dark photon masses. The $(m_{h'},m_{A'})$ plane is scanned 
in steps of 10 MeV in both directions between $0.8 < m_{h'} < 10 \gev$ and $0.25 <m_{A'}<3 \gev$. For each mass hypothesis, 
the signal region is taken as the interval $m_{h'} - 5 \sigma_{m_{h'}} < m_{h'} <m_{h'} + 3 \sigma_{m_{h'}}$ and 
$m_{A'} - 5 \sigma_{m_{A'}} < m_{A'} < m_{A'} + 3 \sigma_{m_{A'}}$, where $\sigma_{m_{A'}}$ ($\sigma_{m_{h'}}$) denotes the corresponding 
dark photon (Higgs) mass resolution. An asymmetric range is used to accommodate the non-Gaussian tail of the low-mass 
side of the signal. The dark photon (Higgs) mass resolution varies between $2-17 \mev$ ($3-55 \mev$), depending on the 
dark photon (Higgs) mass and final state. While setting the limits we adopt the most conservative approach, treating 
as signal every observed event in the signal region. The systematic uncertainties are included by convolving 
the likelihood of each final state with Gaussian distributions having variances equal to the systematic uncertainties 
described below taking correlations into account.

The efficiency is determined for several values of dark photon and Higgs boson masses, and is linearly interpolated 
between the known points. The efficiency includes acceptance, trigger, selection criteria and the dark photon 
branching fraction. The branching fractions into leptons and hadrons are given by 
$BF(A' \rightarrow \ellell) = 1/(2+R)$, $BF(A' \rightarrow {\rm hadrons}) = R/(2+R)$ and 
$BF(A' \rightarrow \pi^+\pi^-) = BF(A' \rightarrow {\rm hadrons}) \sigma(\epem \rightarrow \pi^+\pi^-)/\sigma(\epem \rightarrow {\rm hadrons})$, 
where $R$ denotes the ratio $\sigma(\epem \rightarrow {\rm hadrons}) / \sigma(\epem \rightarrow \mpmm)$ \cite{Bib::pdg}. 
The efficiency increases from a few per mille in regions with small branching fractions to 33\% for the six 
electron mode in the region $m_{A'} < 0.2 \gev$. It drops rapidly in the region $m_{h'} < 0.8 \gev$ and 
$m_{h'} > 10 \gev$, as tracks produced by dark photon decays have a low transverse momentum or are emitted 
close to the beam and are not reconstructed.

\begin{figure}[!htb]
\begin{center}
\includegraphics[width=0.5\textwidth]{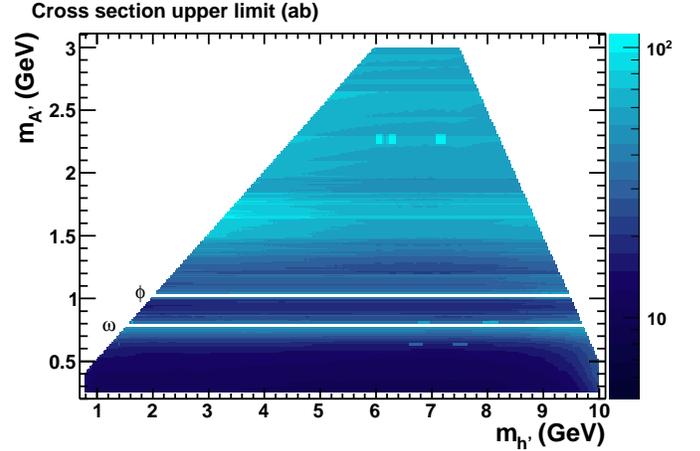}
\caption{Upper limit (90\% CL) on the $\epem \rightarrow A' h', h' \rightarrow A' A'$ cross-section 
as a function of the dark photon and dark Higgs masses. The limits in the $\omega$- and $\phi$-mesons regions are
orders of magnitude larger than the average limits and the corresponding regions (horizontal bands centered around  
$m_{A'} \sim 0.78 \gev$ and $m_{A'} \sim 1.04 \gev$) are masked to avoid overflow.}
\label{figcross}
\end{center}
\end{figure}

The limits on each channel are then combined to extract 90\% CL upper limits on the 
$\epem \rightarrow  A' h', h' \rightarrow A' A'$ cross-section. The results are displayed in Fig.~\ref{figcross}. The 
limits are typically at the level of $10 -100 \, \rm ab$.

The major contribution to the systematic uncertainty arises from the extrapolation procedure used to 
determine the efficiency, which is estimated by comparing the extrapolated value to the nearest known point. 
This uncertainty increases from 1\% to 8\% in some corners of the phase-space. The uncertainty on the 
branching fractions ranges from a few per mille to 4\%. 
The uncertainty due to the modeling of $A' \rightarrow \rm hadron$ decays in 
inclusive modes is estimated by comparing different fragmentation models. This systematic is 
found to be 4\% reflecting the limited sensitivity of the selection procedure to the hadronic 
system produced by the dark photon decay. The uncertainty due to PID algorithms varies between 
1.5\% and 4.5\%, assessed using high-purity samples of leptons and pions.  Additional uncertainties 
include the determination of the track reconstruction efficiency (1.2\%), luminosity (0.6\%), and 
the limited Monte Carlo statistics ($0.5\% - 2.4\%$). 

\begin{figure}[!htb]
\begin{center}
\includegraphics[width=0.5\textwidth]{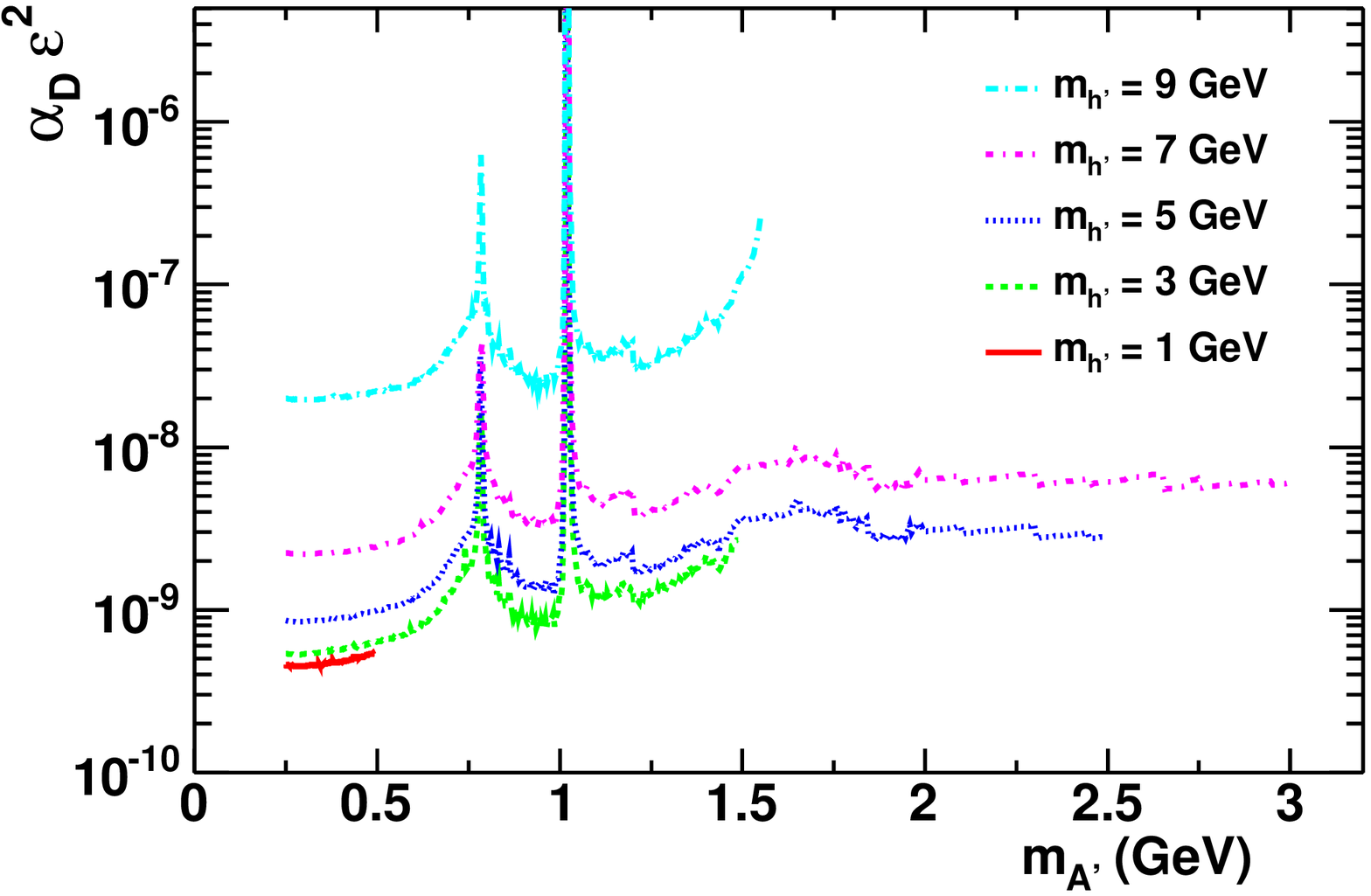}\\
\includegraphics[width=0.5\textwidth]{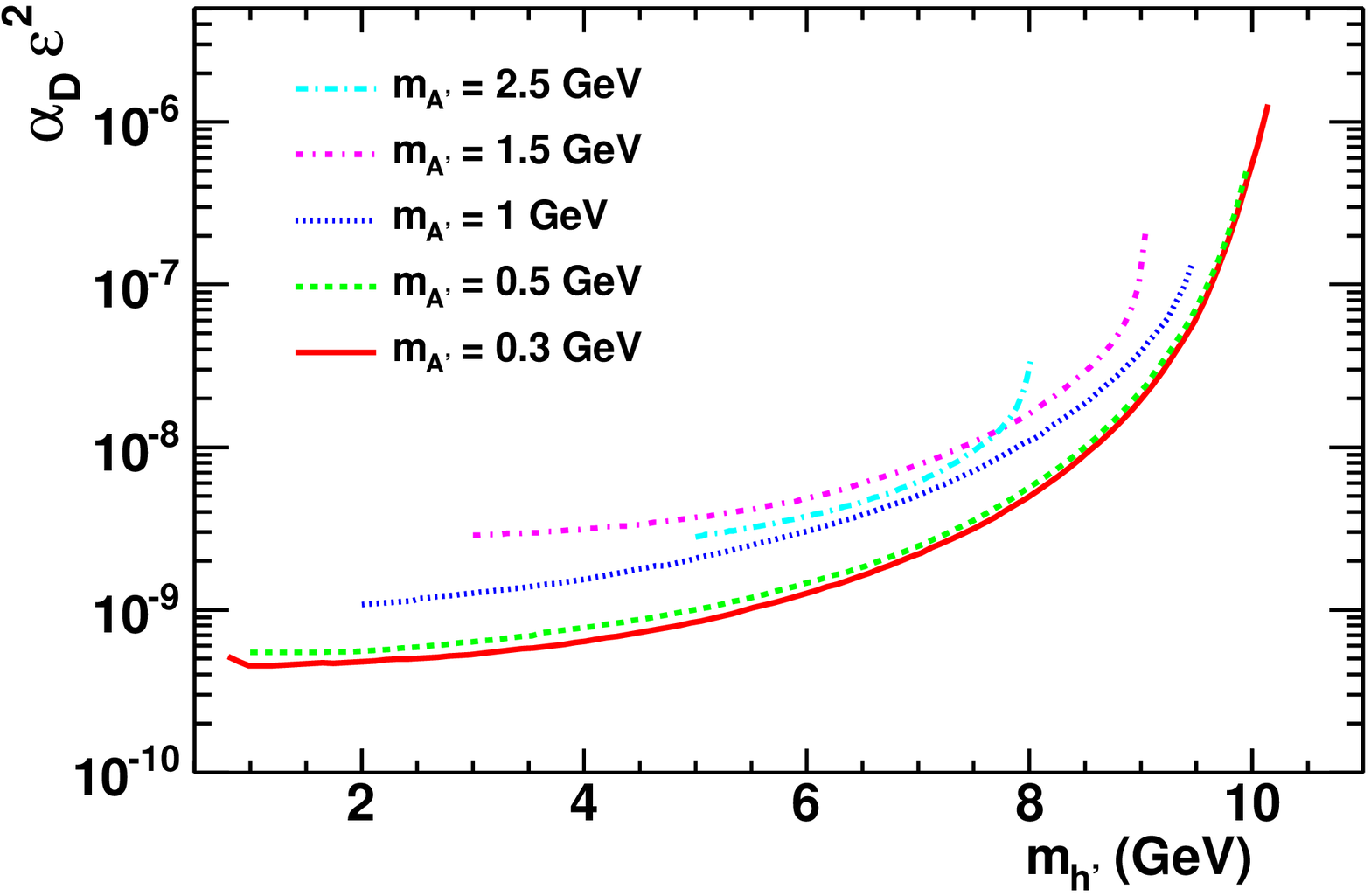}
\caption{Upper limit (90\% CL) on the product $\alpha_D \epsilon^2$ as a function of the dark photon mass for selected values 
of dark Higgs boson masses (top) and as a function of the dark Higgs boson mass for selected values of dark photon
masses (bottom).}
\label{figres}
\end{center}
\end{figure}

The limits on the $\epem \rightarrow  A' h', h' \rightarrow A' A'$ cross section are finally 
translated into 90\% CL upper limits on the product $\alpha_D \epsilon^2$, where 
$\alpha_D = g_D^2/4\pi$ and $g_D$ is the dark sector gauge coupling \cite{Batell:2009yf}. The 
results are displayed in Fig.~\ref{figres} as a function of the dark photon (Higgs) mass for 
selected values of the dark Higgs boson (photon) mass. Values down to $10^{-10} - 10^{-8}$ are 
excluded for a large range of dark photon and dark Higgs masses. These results assume prompt 
dark Higgs boson and dark photon decays. 

In conclusion, a search for dark Higgs boson production has been performed in the range $0.25 < m_{A'} < 3 \gev$ and 
$0.8 < m_{h'} < 10 \gev$ for $m_{h'} > 2 m_{A'}$. No signal has been observed and upper limits on the product of 
the mixing angle and the dark coupling constant in the case of a hidden sector with an Abelian Higgs boson have 
been set at the level of $10^{-10} - 10^{-8}$. Assuming $\alpha_D = \alpha$, these measurements 
translate into limits on the mixing strength in the range $10^{-4} - 10^{-3}$, an order of magnitude smaller 
than the current bounds. 

\section{Acknowledgments}
\label{sec:Acknowledgments}
The authors wish to thank R.~Essig, N.~Toro and P.~Schuster for useful discussions on theoretical issues.
We are grateful for the excellent luminosity and machine conditions
provided by our \pep2\ colleagues, 
and for the substantial dedicated effort from
the computing organizations that support \babar.
The collaborating institutions wish to thank 
SLAC for its support and kind hospitality. 
This work is supported by
DOE
and NSF (USA),
NSERC (Canada),
CEA and
CNRS-IN2P3
(France),
BMBF and DFG
(Germany),
INFN (Italy),
FOM (The Netherlands),
NFR (Norway),
MES (Russia),
MICIIN (Spain),
STFC (United Kingdom). 
Individuals have received support from the
Marie Curie EIF (European Union),
the A.~P.~Sloan Foundation (USA)
and the Binational Science Foundation (USA-Israel).

\end{document}